\documentclass[11pt]{iopart}
\usepackage{graphicx}
\usepackage{color}

\begin{document}
\title[Distorted magnetic orders and electronic structures of tetragonal FeSe]{Distorted
magnetic orders and electronic structures of tetragonal FeSe from
first-principles}
\author{Yong-Feng Li, Li-Fang Zhu, San-Dong Guo, Ye-Chuan Xu, and Bang-Gui Liu}
\address{Institute of Physics, Chinese Academy of Sciences,
Beijing 100190, China} \address{Beijing National Laboratory for
Condensed Matter Physics, Beijing 100190, China}
\date{\today}

\ead{bgliu@mail.iphy.ac.cn}

\begin{abstract}
We use the state-of-the-arts density-functional-theory method to
study various magnetic orders and their effects on the electronic
structures of the FeSe. Our calculated results show that, for the
spins of the single Fe layer, the striped antiferromagnetic orders
with distortion are more favorable in total energy than the
checkerboard antiferromagnetic orders with tetragonal symmetry,
which is consistent with known experimental data, and the
inter-layer magnetic interaction is very weak. We investigate the
electronic structures and magnetic property of the distorted phases.
We also present our calculated spin coupling constants and discuss
the reduction of the Fe magnetic moment by quantum many-body
effects. These results are useful to understand the structural,
magnetic, and electronic properties of FeSe, and may have some
helpful implications to other FeAs-based materials.
\end{abstract}


\maketitle

\section{Introduction}

The advent of superconducting F-doped LaFeAsO stimulates a
world-wide campaign for more and better Fe-based superconductors
\cite{SC-LaOFeAs}. More superconductors were obtained by replacing
La by other lanthanides or partly substituting F for O, and higher
phase-transition temperatures ($T_c$) were achieved in some of
them \cite{SC-LaOFeAs-n,SC-SmOFeAs-n,SC-SmOFeAs-zhao-55K}.
Furthermore, more series of Fe-based superconductors were found,
including BaFe2As2 series and LiFeAs series
\cite{BaFe2As2_38K,SrFe2As2,SrFe2As2-P,LiFeAs}. The highest $T_c$
in these series reaches 55-56 K in the case of doped SmFeAsO
\cite{SC-SmOFeAs-zhao-55K}. Various explorations have been
performed to elucidate their magnetic orders, electronic
electronic structures, superconductivity, and so on
\cite{SC-LaOFeAs-n-MO,LaOFeAs-ES,LaOFeAs-AFM,LaOFeAs-150K,LaOFeAs-ES-SDW,LaOFeAs-OP,LaOFeAs-EC}.
Recently, superconductivity was found even in tetragonal FeSe
samples under high pressure and $\alpha$ FeSe phases with Se
vacancies
\cite{FeSe-27K,FeSe-pressure,FeSe-cg-sc,FeSe-SC,DFT-FeSe}. Very
recently, SrFeAsF was made superconducting by La and Co doping
\cite{hosono,wen2,epl,wen1,zlf}. The FeSe system is interesting
because its FeSe layer is similar to the FeAs layer of the
FeAs-based materials: $R$FeAsO series ($R$: rare earth elements),
$A$Fe2As2 series ($A$: alkaline-earth elements), LiFeAs series,
and SrFeAsF series. In addition to the FeAs layers, there are $R$O
layers in $R$FeAsO series, $A$ layers in $A$Fe2As2 series, Li
layers in LiFeAs series, and SrF layers in SrFeAsF series, but
there is nothing else besides the FeSe layers for the FeSe
systems. Therefore, it is highly desirable to investigate the
magnetic orders, electronic structures, and magnetic properties of
the tetragonal FeSe phases (or distorted phases of them).

In this article we use an full-potential density-functional-theory
method to study various magnetic orders and their effects on the
electronic structures of the FeSe. We suppose checkerboard
antiferromagnetic order for the spins of the Fe layer of the
tetragonal phase and striped antiferromagnetic orders for those of
the symmetry-broken structures, and perform total-energy and force
optimization to determine the structural and magnetic parameters.
Then, we investigate the corresponding electronic structures and
magnetic property of them. Our calculated result means that the
striped antiferromagnetic order is favorable for the spins of the
Fe layer, which is consistent with main known experimental data.
We also discuss the reduction of the Fe magnetic moment by quantum
many-body effects. More detailed results are presented in the
following.

The paper is organized as follows. In next section, we give our
computational detail. In the third section, we present our main
DFT calculated results, including optimized magnetic structures
and corresponding electronic density of states and energy bands.
In the fourth section, we discuss spin interactions and many-body
effects on the Fe magnetic moments. Finally, we present our main
conclusion in the fifth section.

\section{Computational detail}

Our calculations are performed by using a full-potential
linearized augmented plane wave (FLAPW) method within the density
functional theory (DFT)\cite{dft}, as implemented in the Vienna
package WIEN2k \cite{wien2k}. The generalized gradient
approximation (GGA) is used for the exchange and correlation
potentials \cite{pbe96}. We take the 3$d$ and 4$s$ states of Fe
and the 4$s$ and 4$p$ states of Se as valence states, and the 3$p$
states of Fe and 3$d$ states of Se are treated as semicore states.
The core states include all the lower states. The core states are
treated in terms of radial Dirac equation and thus the full
relativistic effect is included. For the valence and semicore
states, the relativistic effect is calculated under the scalar
approximation, with spin-orbit interaction being neglected
\cite{relsa}. The radii of Fe and Se muffin-tin spheres are 2.05
and 2.00 a.u., respectively. To get more accurate results, we take
$R_{mt}\times K_{max}$=9.0 and make the angular expansion up to
l=10 in the muffin-tin spheres. We use 1000 k points in the
calculations. For different magnetic orders the k points in the
first Brillouin zone are chosen differently because of the
different symmetry. The self-consistent calculations are
controlled by the charge density, and the convergence standard is
that the difference between input charge density and output one is
less than 0.00005 per unit cell.

\begin{figure}[tb]
\begin{center}
\scalebox{0.9}{\includegraphics{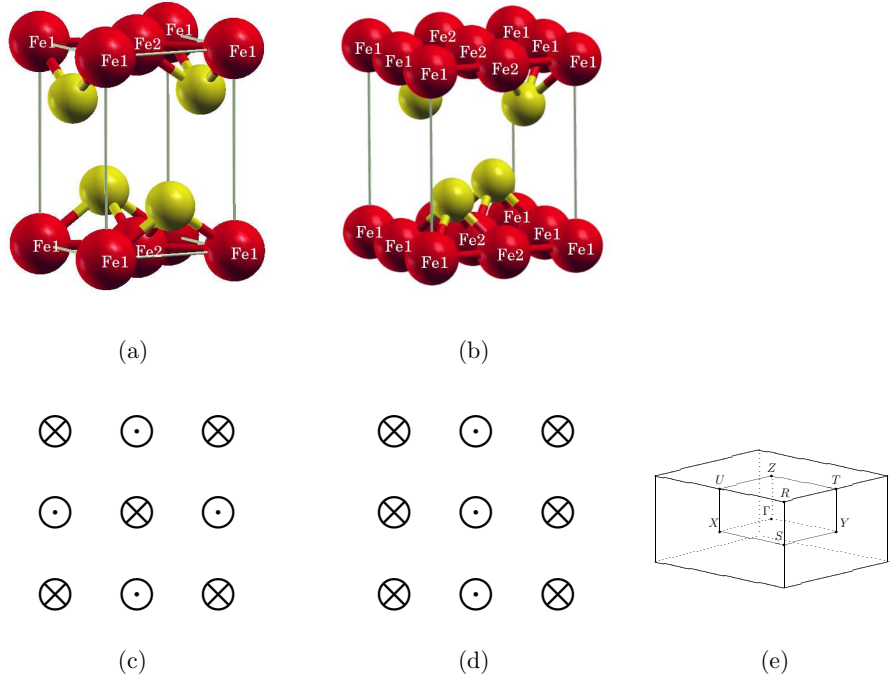}} \caption{(color online).
Schematic structures of tetragonal FeSe (a) and orthorhombic FeSe
(b); and the spin configurations of the checkerboard
antiferromagnetic order (c) and the striped antiferromagnetic order
(d). The big balls with Fe1 and Fe2 indicates the two kinds of Fe
atoms with different spin orientations, and the small one Se atom.
The sign $\otimes$ means that the Fe1 spin orients upward with
respect to the paper plane, and $\odot$ the Fe2 spin downward. The
first Brillouin Zone of the orthorhombic FeSe is shown in (e) with
the high-symmetry points labelled.}
\end{center}
\end{figure}

We show the unit cell of the tetragonal FeSe (PbO structure) in
Fig. 1a. It is a layered structure, in which Fe atom occupies 2a
position and Se atom 2c position. There are six possible magnetic
configurations for the Fe ions if the tetrahedral FeSe structure
is allowed to distort into orthorhombic structures. Fig. 1e shows
the first Brillouin zone with the representative points and lines.
The Fe moments in the Fe plane can be arranged to form
ferromagnetic or antiferromagnetic orders. The AFM orders can be
checkerboard-like (Fig. 1c) or striped (Fig. 1d). For the
successive \{001\}planes, the Fe moments can couple
ferromagnetically or antiferromagnetically. As a result, we have
four different antiferromagnets, namely (a) checkerboard-FM, (b)
checkerboard-AF, (c) stripe-FM, and (d) stripe-AF. In addition,
when we force Fe moments in the Fe plane have FM order, the
self-consistent calculations yield zero moments for the Fe
moments, which means that FM order is unstable for FeSe,
independent of the interlayer spin arrangements. Therefore, we can
actually construct five magnetic orders for this system.

\section{Main calculated results}

\begin{table}[b]
\begin{center}
\caption{The magnetic order, the relative total energy per formula
unit ($\Delta E$ in meV, with the lowest stripe-FM set as
reference), the magnetic moment in the Fe muffin-tin sphere ($M$
in $\mu_B$), and the internal Se position parameter $u_{\rm Se}$
of the two striped antiferromagnetic orders and the two
checkerboard ones. The results of the nonmagnetic order are
presented for comparison. }
\begin{tabular}{ccccccccc}
 \hline\hline
& Magnetic order & & $\Delta E$ (meV) & & Moment $M$ ($\mu_{B}$) & & $u_{\rm Se}$ &\\
\hline
& checkerboard-FM & & 72 & & 1.82 & & 0.2570 &\\
& checkerboard-AF & & 72 & & 1.81 & & 0.2426 &\\
\hline
& stripe-FM & & 0 & & 1.98 & & 0.2590 &\\
& stripe-AF & & 5 & & 2.00 & & 0.2592 &\\
\hline
& nonmagnetic & & 154 & & 0.00 & & 0.2471 &\\
 \hline\hline
\end{tabular}
\end{center}
\end{table}

\begin{figure}[!htb]
\begin{center}
\scalebox{0.8}{\includegraphics{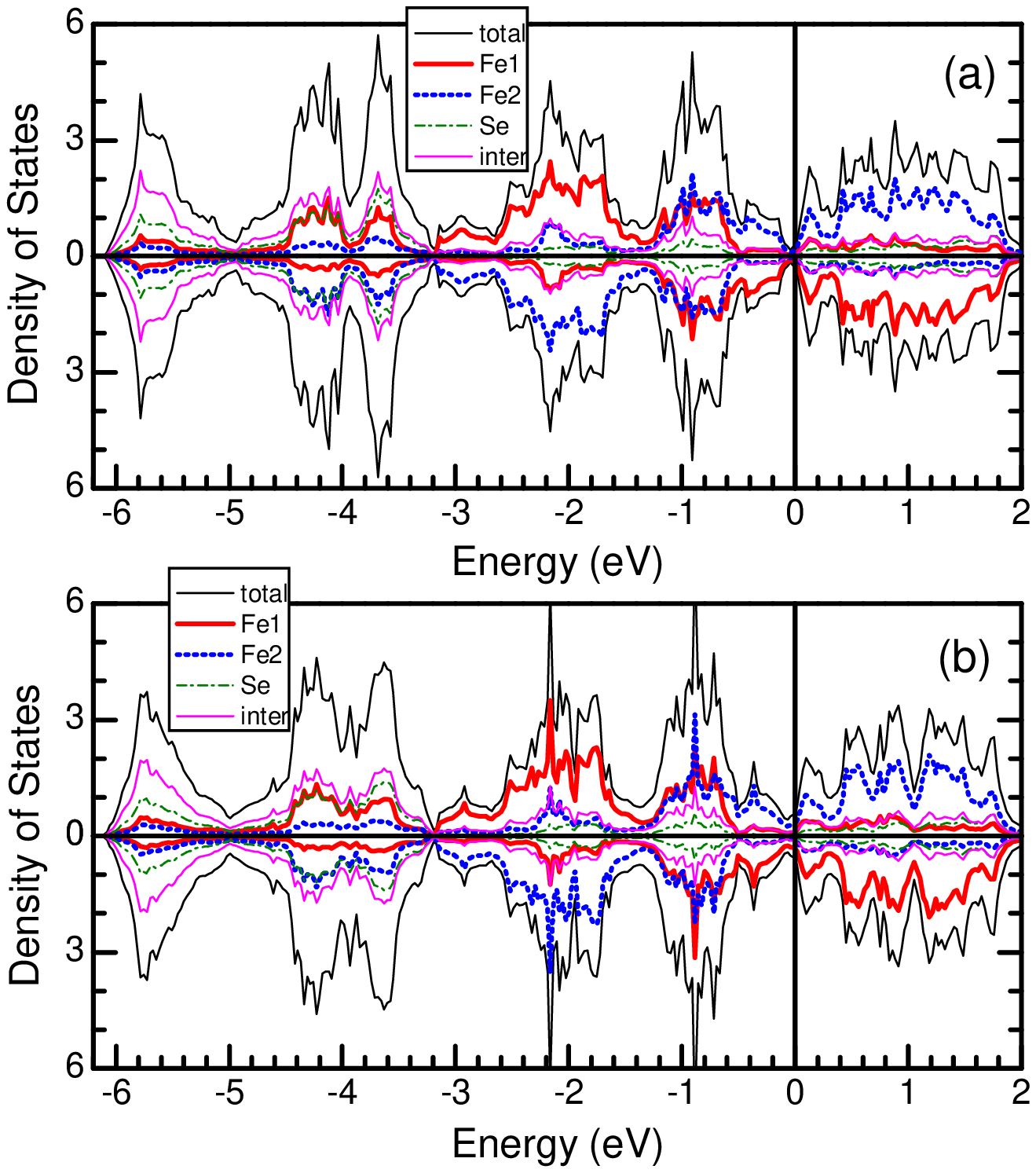}} \caption{(color
online). Spin-dependent density of states (DOS, in units of
states/eV per formula unit) of the stripe-FM (a) and stripe-AF
(b). The upper part of each panel is the majority-spin DOS and the
lower part the minority-spin DOS. The Fe1 DOS is emphasized by
thick red (or gray) solid lines, and the Fe2 DOS by thick blue (or
gray) dotted lines. The black thin solid line indicates the
spin-dependent total DOS, and the others are projected DOS in the
muffin tin sphere of Se atom (dot-dash) and the interstitial
region (pink or gray thin solid). }
\end{center}
\end{figure}

The parameters in our calculations are taken from the experimental
values. We use $a=3.76 \mathrm{\AA}$ and $c=5.52\mathrm{\AA}$. The
positions of Se atoms are optimized fully, and the force of Se
atom is made less than 2 mRy/a.u. Calculated results of total
energy, magnetic moment, and Se position parameter are summarized
in Table I. It makes little difference in total energy to arrange
interlayer Fe moments in FM or AF order. The total energy results
reveal that the intralayer Fe-Fe interaction is strong and the
interlayer interaction weak. Actually, the striped arrangement of
the Fe moments lowers the total energy of the FeSe layer
approximately by 70 meV with respect to the checkerboard
arrangement. On the other hand, the non-magnetic order is 154 meV
higher than the lowest striped AF order. Therefore, the striped AF
spin order is the magnetic ground state in the FeSe layer, but the
actual interlayer magnetic interaction is too weak for any
density-functional-theory calculation to determine. The moment in
the Fe muffin-tin sphere is about 1.8$\mu_B$ for the two
checkerboard orders, and about 2.0$\mu_B$ for the two striped
orders. The total moment for one Fe atom is estimated to be a
little larger for the striped AF orders. For the striped AF
orders, the Se position parameter remains almost the same when the
interlayer spin coupling is changed from FM to AF. Because it is
impossible to distinguish between FM and AF spin alignment in the
$z$ direction by density-functional-theory calculation, we present
calculated results for both FM and AF arrangement in the $z$
direction in the following.

\begin{figure}[!htb]
\begin{center}
\scalebox{1.1}{\includegraphics{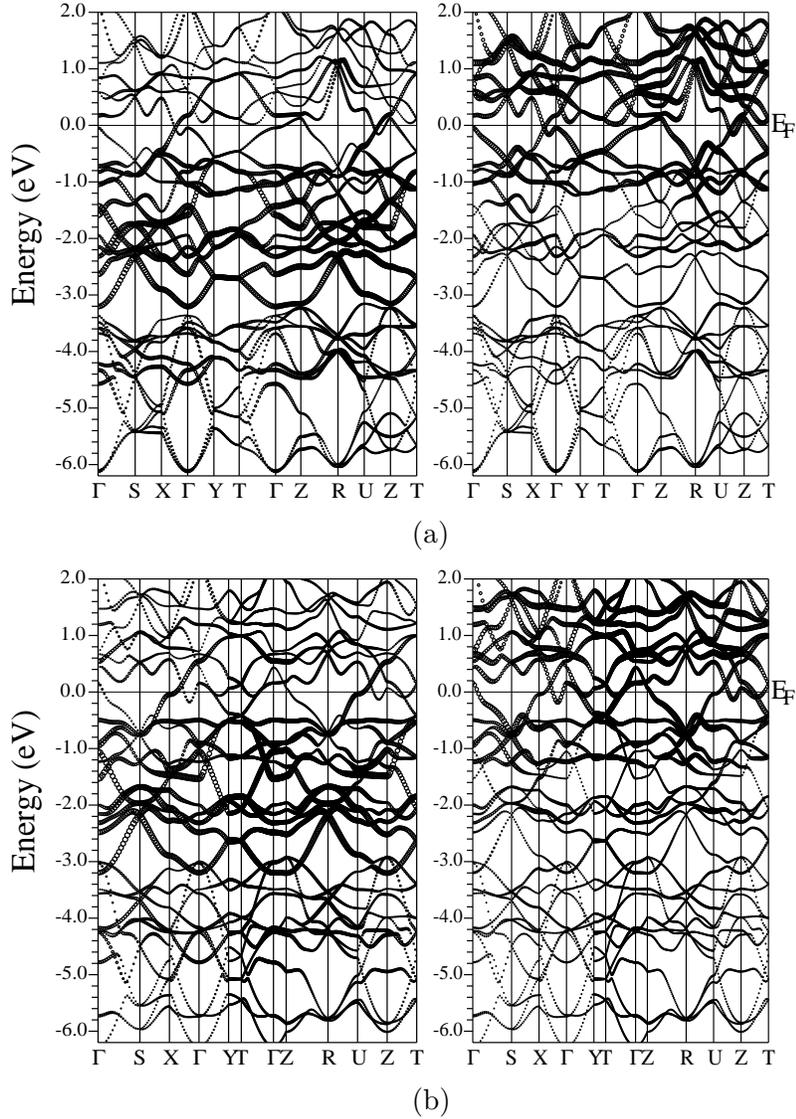}} \caption{Spin-dependent
energy bands of the stripe-FM (a), and stripe-AF (b). The left
panel for each magnetic order shows the energy bands of Fe1
spin-up channel with the Fe1 atomic character emphasized by
circles, and the right one those of Fe1 spin-down channel. The
spin of Fe1 orients in antiparallel to that of Fe2. The emphasized
energy bands consist of circles with different diameters, and the
larger the diameter, the more the atomic character. There are so
many circles (or dots) for a given band that they seams to be
connected forming a line for the band.}\end{center}
\end{figure}

We present the spin-dependent density-of states (DOS) of the FeSe
in the two striped AF orders in Fig. 2. There is no energy gap
near the Fermi levels and therefore the FeSe for each AF order
shows a metal feature. The Fe atom has different crystalline
environment for different magnetic orders, and thus its states are
reformed in different ways. It can be seen that the main peaks
occupy the states in the energy window from -2.2 eV to -1.5 eV.
Almost all the partial DOS of Fe atom comes from the 3d sates, and
the DOS of Se atom consists mainly of the p states.

We present the electronic energy bands of the FeSe in the two
striped AF orders in Fig. 3. The plots look like lines, but
consist of hollow circles. We choose the $k$ points uniformly for
convenient comparison. The circle diameter is proportional to the
Fe1 spin-up or Fe1 spin-down atomic character of the band at that
$k$ point. The dispersion along the $z$ direction is much stronger
than those in other similar Fe-based materials. This can be
attributed to the short distance between the successive Fe layers.
It can be seen that the d states of Fe play a key role for all the
magnetic structures. For the stripe magnetic structures we can
find the electrons and holes along the X-$\mathrm{\Gamma}$-Y line,
which is very important to achieving superconductivity in doped
cases or under appropriate pressures.

\section{Spin interactions and many-body effects on the magnetic moments}

In order to further investigate the magnetic moment, we use the
following AFM Heisenberg spin model to describe the spin
properties of the Fe atoms in the striped AFM phase.
\begin{equation}
 H=\sum_{ij}J_{ij}\vec{S}_i\cdot \vec{S}_j
\end{equation}
where $\vec{S}_i$ is quantum spin operator for site $i$, and
$J_{ij}$ is the exchange coupling constants between the two spins
at sites $i$ and $j$. For the striped AFM phase, the nearest
coupling constant in the $x$ direction is $J_x$, and that in the
$y$ direction $J_y$. For the tetrahedral phase, we should have
$J_x=J_y$, but for the striped phase we have $J_x\ne J_y$ because
the crystalline distortion in the $xy$ plane. We limit non-zero
exchange coupling constants up to the next nearest neighboring
spins, $J^\prime$. For convenience, we define $J=(J_x+J_y)/2$ and
$\delta=J_x-J_y$. Our DFT calculation yields $J_x=10$meV,
$J_y=8$meV, and $J^\prime=5$meV. These means $J=9$meV and
$\delta=2$meV. The $J_x$ and $J_y$ comes from the superexchange
through the two nearest Se atoms and $J^\prime$ from that through
the one nearest Se atom, and as a result, we should have $J\approx
2 J^\prime$, which supports our DFT results.

\begin{figure}[!htb]
\begin{center}
\scalebox{0.7}{\includegraphics{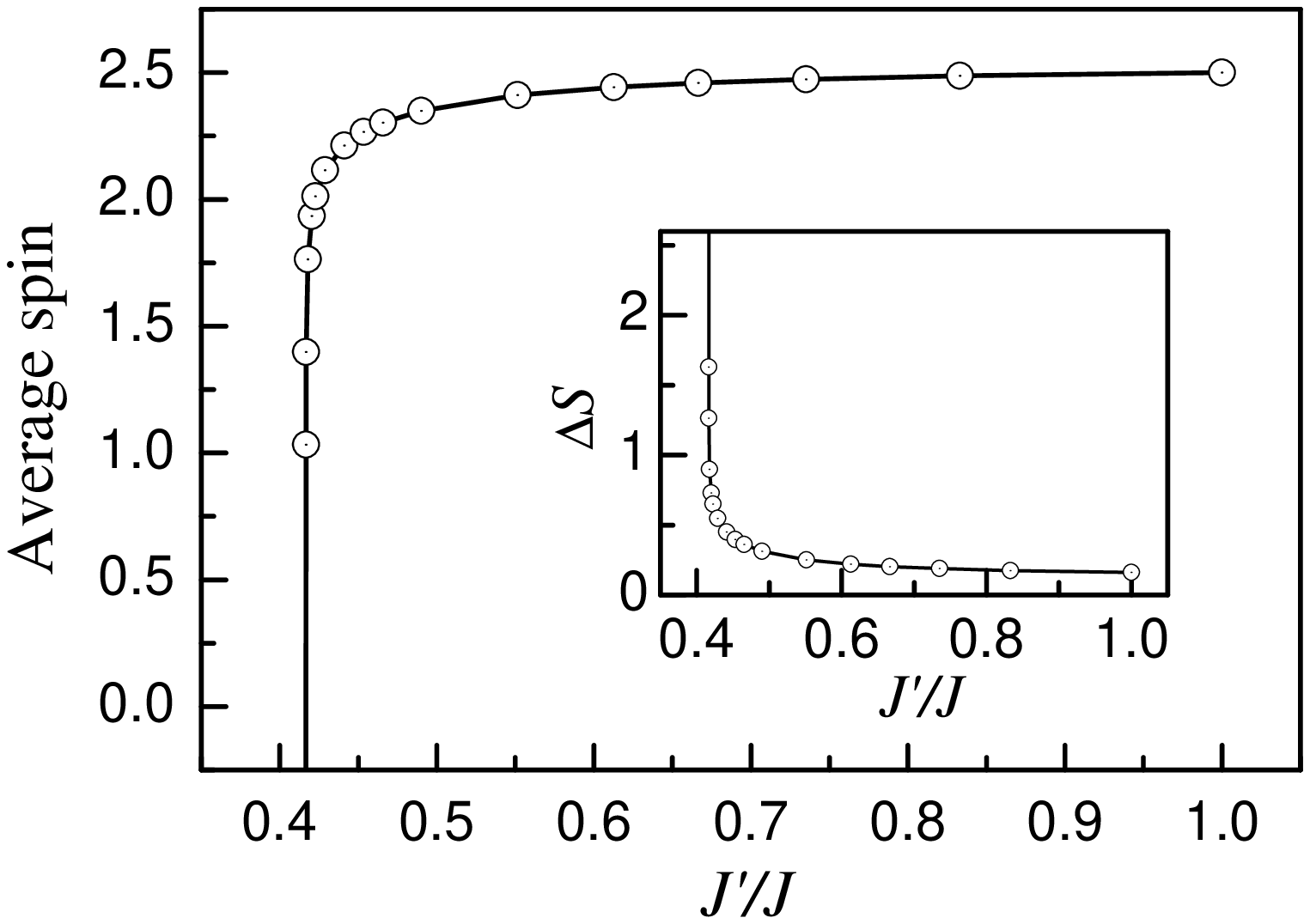}}
\caption{Zero-temperature average
 spin $\langle S^z\rangle$ vs. the coupling
constant ratio $J^\prime/J$ for the two-dimensional spin model.
The inset shows $\Delta S$ (defined in text) vs. $J^\prime/J$. The
$\odot$ signs indicate calculated points.}
\end{center}
\end{figure}

We treat the spin Hamiltonian (1) with the above parameters with
spin wave theory\cite{lbg}. As usual, the average spin for zero
temperature, $\langle S^z\rangle$, can be given by $\langle
S^z\rangle=S-\Delta S$, where $S$ is the spin value of the Fe atom
in the FeSe and $\Delta S$ the correction due to quantum many-body
effects. Presented in Fig. 4 are our calculated results for
$\langle S^z\rangle$ and $\Delta S$ as functions of the parameter
$J^\prime/J$. It is clear that $\langle S^z\rangle$ decreases with
decreasing $J^\prime/J$, getting to zero at $J^\prime=J_y/2$. In
fact, the striped AFM structure is not the magnetic ground state
of the FeSe any more if $J^\prime$ is smaller than $J_y/2$. For
real samples of the FeSe, one should have the parameter relations
$J^\prime \approx J/2$ and $\delta$ is small but finite, and
therefore the experimental spin value is small compared to the DFT
value $S$.

\section{Conclusion}

In summary, we have used the full-potential
density-functional-theory method to study various magnetic orders
and their effects on the electronic structures of the FeSe. We
find that, for the spins of the single Fe layer, the striped
antiferromagnetic orders with the broken symmetry are more
favorable in total energy than the checkerboard antiferromagnetic
orders with tetragonal symmetry, and the inter-layer magnetic
interaction is very weak. Then, we investigate the corresponding
electronic structures and magnetic property of the distorted
phases with the striped antiferromagnetic orders. Our calculated
result that the striped antiferromagnetic order is favorable for
the spins of the Fe layer is consistent with main known
experimental data. We also present our calculated spin coupling
constants and conclude that the reduction of the Fe magnetic
moment is caused by quantum many-body effects. These results are
useful to understand the structural, magnetic, and electronic
properties of FeSe, and may have some helpful implications to
other FeAs-based materials.

\section*{Acknowledgements}

This work is supported  by Nature Science Foundation of China
(Grant Nos. 10774180, 10874232, and 60621091), by Chinese
Department of Science and Technology (Grant No. 2005CB623602), and
by the Chinese Academy of Sciences (Grant No. KJCX2.YW.W09-5).

\end{document}